\documentclass[twocolumn, aps, prl, unsortedaddress]{revtex4-1}
\usepackage{amsmath,graphicx,amssymb}

\begin{document}

\title{Microscopic and macroscopic signatures of 3D Anderson localization of light}

\author{Florent Cottier}
\affiliation{Instituto de F\'{i}sica de S\~{a}o Carlos, Universidade de S\~{a}o Paulo - 13560-970 S\~{a}o Carlos, SP, Brazil}
\affiliation{Universit\'{e} C\^{o}te d'Azur, CNRS, INPHYNI, France}

\author{Ana Cipris}
\affiliation{Universit\'{e} C\^{o}te d'Azur, CNRS, INPHYNI, France}

\author{Romain Bachelard}
\affiliation{Departamento de F\'{\i}sica, Universidade Federal de S\~{a}o Carlos, Rod. Washington Lu\'{\i}s, km 235 - SP-310, 13565-905 S\~{a}o Carlos, SP, Brazil}
\affiliation{Universit\'{e} C\^{o}te d'Azur, CNRS, INPHYNI, France}

\author{Robin Kaiser}
\affiliation{Universit\'{e} C\^{o}te d'Azur, CNRS, INPHYNI, France}

\date{\today}


\begin{abstract}
Apart from the difficulty of producing highly scattering samples, a major challenge in the observation of Anderson localization of 3D light is identifying an unambiguous signature of the phase transition in experimentally feasible situations. In this letter we establish a clear correspondence between the collapse of the conductance, the increase in intensity fluctuations at the localization transition and the scaling analysis results based on the Thouless number, thus connecting the macroscopic and microscopic approaches of localization. Furthermore, the transition thus inferred is fully compatible both with the results based on the eigenvalue analysis of the microscopic description and with the effective-medium Ioffe-Regel criterion.
\end{abstract}

\maketitle

After the pioneering work of Ph.~W.~Anderson on localization of electron waves in metals due to disorder~\cite{Anderson1958}, where the electronic modes acquire an exponentially localized profile in the crystalline structure, the phenomenon was shown to hold more generally for propagation of waves in disordered potentials, including light~\cite{John1983,John1987} and acoustic~\cite{He1986} waves. Since then, efforts focused on observing localization induced by disorder, independently of the interparticle interactions present for electrons: experimental signatures of Anderson localization have been reported for acoustic waves~\cite{Hu2008}, matter waves in one and three dimensions~\cite{Billy2008,Roati2008,Kondov2011,Semeghini2015}, micro-waves in quasi-1D~\cite{Chabanov2000} and two dimensions~\cite{Laurent2007}, and even surface plasmon polaritons~\cite{Shi2018}.

Localization of light has a more tortuous story~\cite{Wiersma2013,Segev2013}, due to the lack of definite signatures and of highly scattering systems. Indeed while localization has been reported for propagating light waves in 1D~\cite{Berry1997} and 2D~\cite{Laurent2007}, these lower dimensional systems enable a direct observation of the modes, and do not suffer the necessity of a critical disorder strength to reach localization. Differently, 3D samples are difficult to produce, so it is difficult to assess a subwavelength scattering mean free path~\cite{Skipetrov2018}, and the initial signatures for 3D experiments, looking at either the diffuse transmission~ \cite{Genack1991,Wiersma1997,Schuurmans1999,Sperling2012} or at the late time decay of the transmitted intensity~\cite{Strzer2006}, were later challenged. 
On the one hand, high scattering cross-section samples were obtained from semi-conductor powders, yet they were plagued with absorption~\cite{Scheffold1999}, masking or mimicking localization.
On the other hand, TiO$_2$ compressed to high densities, used in time-resolved experiments, required high laser input power to detect possible deviations from diffuse transmission laws. This leads to nonlinear effects and spurious fluorescence of the sample, and the corresponding long lifetimes could again be incorrectly interpreted as localization~\cite{Sperling2016}. In addition to these technical challenges,  experimental observables which would exhibit a smoking gun signature of localization are hard to obtain in 3D. Indeed, it is difficult if not impossible to perform a direct observation of localized eigenstates, and one is often limited to detecting light scattered by the sample.

These controversial results originate in gaps in the localization theory of light, where microscopic and macroscopic approaches have not been unified yet. On the one hand, important numerical efforts were recently dedicated to elucidate the crucial role of polarization and near-field terms in precluding localization~\cite{Skipetrov2014, Bellando2014}, or the possibility of restoring it using an external magnetic field~\cite{Skipetrov2015,Skipetrov2018b}. These decisive results were obtained through eigenvalue analysis (the so-called scaling analysis~\cite{Abrahams1979}) of a microscopic representation of cold atom clouds~\cite{Lehmberg1970}. On the other hand, the definition of a macroscopic conductance  from the light scatistics aims at describing the modification of transport at the localization transition~\cite{vanRossum1999}, but a formal connection to the microscopically derived quantities, including a microscopic conductance derived from eigenvalues~\cite{Skipetrov2014, Bellando2014, Skipetrov2018}, is still missing. Since neither eigenvalues nor (localized) eigenvectors are easily measured in optical experiments, the important question of a proper macroscopic observable to detect the localization transition of light remains unanswered.

In this Letter, starting from an ab initio description of the light-atom interaction we show that intensity statistics allow to connect the microscopic and macroscopic realms, and are a suitable observable for the study of the localization transition. 
Intensity fluctuations have been used to detect localization of ultrasound waves in 3D~\cite{Hu2008} and of light in quasi-1D~\cite{Chabanov2000}. We here show that the statistics (analyzed through the intensity variance) present deviations from Rayleigh law in a one-to-one correspondence with the macroscopic conductance, which collapses at the phase transition.
 The extension of this study in presence of a magnetic field gives results in excellent agreement with the eigenvalue analysis, as well as with the effective medium Ioffe-Regel criterion on the scattering mean free path, showing the full compatibility between the different approaches. This suggests that intensity statistics are a powerful tool to study the 3D localization transition for light.

\begin{figure}[!h]
\centering
\includegraphics[width=0.5\textwidth]{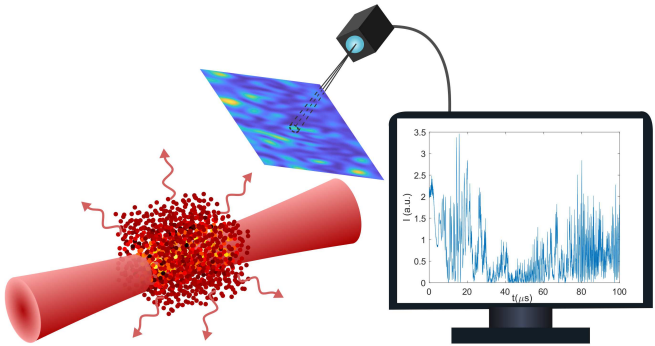} 
\caption{Scheme to detect fluctuations in the radiated intensity by a cold atomic cloud. The inset presents a time signal obtained from simulations realized with Eq.\eqref{Eq:CDStationary}, for atoms at a finite temperature.\label{Fig:setup}}
\end{figure}

An experimental scheme to measure intensity fluctuations of the scattered light is presented in Fig.\ref{Fig:setup}, where the cloud is illuminated by a focused laser beam, and its radiation is collected in a given direction. The inset shows the result of a numerical simulation of the fluctuating detected intensity, due to the motion of the atoms trapped at finite temperature in a harmonic potential. The fluctuations of radiated intensity is captured by the second order optical coherence $g^{(2)}(\tau) = \langle I(t)I(t-\tau)\rangle/\langle I^2(t)\rangle$, where $\langle.\rangle$ refers to a time average. Throughout this letter, the fluctuations analysis is performed using various distributions of motionless atoms (considering the stationary regime for the dipoles of Eq.\eqref{Eq:CDStationary} below), so the average $\langle.\rangle$ hereafter refers to a configuration average. It allows to explore larger systems at a much lower computational cost, and we have checked that both approaches lead to the same conclusions.

To address the light-matter interaction from a microscopic point of view, we use an ab initio model of two-level systems.
The atomic cloud is modelled as an ensemble of $N \gg 1$  point scatterers randomly distributed in a cube of side length $L$ with a uniform density $\rho = N/L^3$, with fixed positions $\mathbf{r}_j$. The two-level atoms present a transition at frequency $\omega_a$, of linewidth $\Gamma$, and they are driven with a monochromatic Gaussian beam of waist $w_0 = L/4$ and wavevector $\mathbf{k}_L = k\mathbf{\hat{z}}$, detuned from the transition by $\Delta = \omega - \omega_a$, with $\Delta\ll\omega_a$ so one can assume $k\approx\omega_a/c$. We consider a low Rabi frequency $\Omega_L(\mathbf{r})=dE_L(\mathbf{r})/\hbar\ll\Gamma$, with $d$ the electric dipole moment, so that the scattering process is elastic. Using the Markov and rotating wave approximations, and neglecting at first polarization effects, the dynamics of the atomic dipoles $\beta_j$ is given by a set of $N$ coupled equations~\cite{Javanainen1999,Svidzinsky2010}:
\begin{eqnarray}\nonumber
\frac{d\beta_j}{dt} &=& \left( i\Delta-\frac{\Gamma}{2} \right) \beta_j - \frac{id}{2\hbar} E_L(\mathbf{r}_j) 
\\ &&- \frac{\Gamma}{2} \sum_{m \neq j} \frac{\exp(i k|\mathbf{r}_j-\mathbf{r}_m|)}{i k|\mathbf{r}_j-\mathbf{r}_m|}\beta_m,\label{Eq:CDStationary}
\end{eqnarray}
where the last term describes the effective dipole-dipole interaction. 
In the far--field limit, the radiated intensity at a point $r\hat{\mathbf{n}}$ reads:
\begin{equation}
I(\hat{\mathbf{n}})\propto\bigg|\frac{id}{\hbar\Gamma}E_L(\mathbf{r})+\frac{e^{ikr}}{ikr}\sum_{j = 1}^N e^{-i k\hat{\mathbf{n}}.\mathbf{r}_j}\beta_j\bigg|^2.
\end{equation}
Whereas in the diffusive regime the average intensity scattered around the forward direction
decreases with the sample size $L$ as $1/L$ (known as Ohm's law for photons), it is expected to present an exponential decay with the system length in the localized regime. Although demonstrated experimentally in 1D with transparent plates~\cite{Berry1997}, no such observation has been reported in higher dimensions.
A theoretical work  suggested that the late-time radiation by cold atoms may present different characteristics in the localized and the non-localized regimes~\cite{Skipetrov2016b}, yet the role of subradiance was not discarded~\cite{Guerin2016,Weiss2018,Cottier2018}, which would require a systematic study of scaling law of sample size, atom number and laser detuning. So far, our numerical studies on configuration averaged time decay curves of the transmitted intensity around the forward direction using the coupled dipole model did not allow us to identify an unambiguous signature of Anderson localization. 

\begin{figure}
    \centering
    \includegraphics[width=8cm]{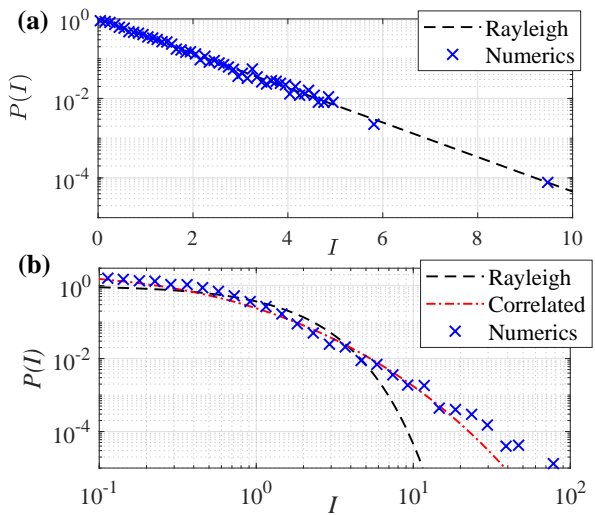}
    \caption{\label{Fig:Histograms}Probability distribution function of the intensity in (a) the dilute case ($\rho=5/\lambda^3$, $g_C\gg 20$) and (b) the localized regime ($\rho=44/\lambda^3$, $ g_C = 0.27\pm0.02$). The black dashed curves refer to Rayleigh law, the crosses were obtained from numerical simulations of Eq.\eqref{Eq:CDStationary} and the red dash-dotted curve was computed from Eq.~\eqref{Eq:PDFGeneral}. Intensity radiated in the angle $(\theta, \varphi)=(5\pi/12,0)$ from the laser axis, from a system of size $kL=32.4$ and (a) $N=684$ or (b) $N=6066$, using $10^4$ realizations.}
\end{figure}

We now turn our attention to the statistics of the scattered intensity around forward direction, as the localization phase transition is expected to present a strong increase in fluctuations~\cite{Hu2008}. As we will show, this approach enables us to identify a clear signature of Anderson localization without the need for finite size scaling or similar demanding numerical simulations. 
In the very dilute limit, multiple scattering and collective effects are unable to correlate the atomic dipoles, and the resulting speckle has a probability distribution function that obeys Rayleigh law~\cite{Goodman1985}: $P(I) = e^{-I}$, where $I$ hereafter refers to the normalized intensity ($I/\langle I\rangle\rightarrow I$): hence, its variance $\sigma_I^2$ is equal to one. This behaviour is illustrated in Fig.~\ref{Fig:Histograms}(a) for a cloud with a low density, below the localization threshold ($\rho<\rho_c\approx 22/\lambda^3$). 

\begin{figure}
    \centering
    \includegraphics[width=0.45\textwidth]{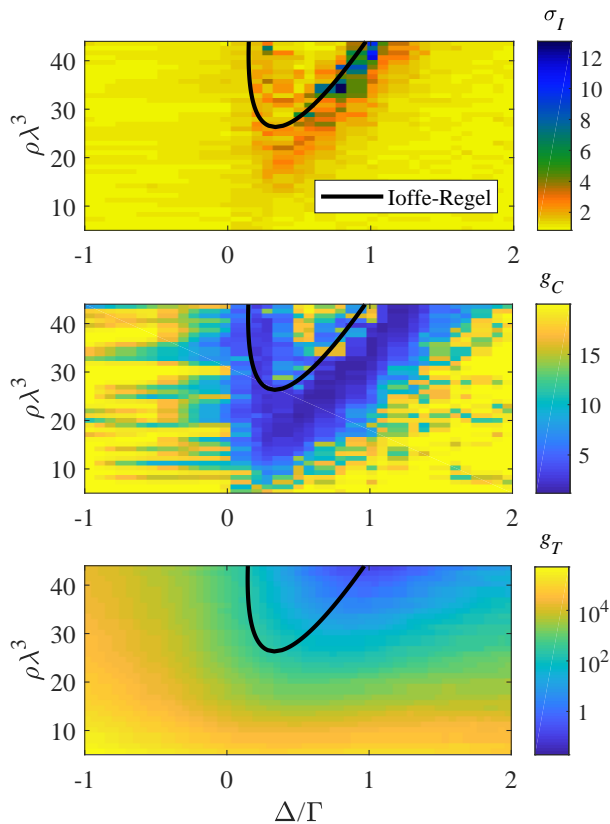} 
    \caption{\label{Fig:condvar}Phase diagram for the (a) intensity variance, (b) conductance $g_C$ and (c) Thouless number $g_T$ in the $(\Delta/\Gamma,\rho\lambda^3)$ plane. Simulations realized for a homogeneous cubic cloud of side length $kL = 32$, using $800$ realizations and an observation angle $\theta=5\pi/12$. The value of the conductance $g_C$  is saturated at the arbitrary value of $20$, as it diverges for non-localized samples. The black curve corresponds to Eq.\ref{eq:IR} for $\alpha=0.5$.}
\end{figure}

For increased densities, deviations from Rayleigh law appear, as can be observed in Fig.~\ref{Fig:Histograms}(b) for a density above the transition threshold. Such deviations have been reported in the past for strongly scattering systems~\cite{Garcia1989,Garcia1993}, which lead to theoretical efforts to include correlations between the scatterers into the intensity statistics law~\cite{Shnerb1991,Kogan1993,Kogan1993b}. In particular, von Rossum and Nieuwenhuizen showed that for a Gaussian beam, the intensity statistics are given by~\cite{vanRossum1999}:
\begin{subequations}\label{Eq:PDFGeneral}
\begin{eqnarray}\label{Eq:PDF1}
P(I) &=& \int_{-i \infty}^{i \infty}\frac{dx}{\pi i}K_0(2\sqrt{-xI})\exp\left( -\Phi_c(x) \right),
\\ \Phi_c(x) &=& g_C \int_0^1 \frac{dy}{y} \log \left( \sqrt{1+\frac{xy}{g_C}} + \sqrt{\frac{xy}{g_C}} \right),\label{Eq:PDF2}
\end{eqnarray} 
\end{subequations}
with $K_0$ the modified Bessel function. Here $g$ is a free parameter called the conductance, which value is obtained from fitting the intensity probability distribution function to Eqs.\eqref{Eq:PDFGeneral}. In the case of a dilute sample, i.e. in the absence of Anderson localization, the conductance extracted from the dilute case of Fig.\ref{Fig:Histograms} yields arbitrarily large values, corresponding to the divergent number of optical modes in an infinite system. 
For convenience, we truncate the values shown at $g_C=20$ in Fig.~\ref{Fig:condvar}(a). In the dense regime however small values of $g$, close to unity ($g_C=0.27\pm0.02$), are obtained. We stress that the conductance as defined in ~\cite{vanRossum1999} is related to the number of accessible optical modes, different from the number of eigenvalues as exploited in~\cite{Skipetrov2014, Bellando2014, Skipetrov2018}. The latter, also called the Thouless number, is rather related to the number of atoms used in the simulations, and is hereafter labelled $g_T$. In the context of mesoscopic transport, intensity fluctuations have been associated to short and long range correlations as well as to universal conductance fluctuations, 
with scaling as $1$, $1/g_C$ or $1/g_C^2$ respectively~\cite{Feng1988, Mello1988, Feng1991, Akkermans2004}. Another correlation function called $C_0$ has been studied~\cite{Shapiro1999} and related to averaged local density of states~\cite{Carminati2009}. Yet, while experiments with constrained geometries measured such correlation functions~\cite{Scheffold1998}, the regime of Anderson localization has not been accessible to these experiments.

Despite there does not exist a formal connection between the conductance $g_C$ and the localization transition, it is enlightening to monitor the behaviour of $g_C$ in the range of densities and energies known to exhibit the localization transition~\cite{Skipetrov2018}. We compare it directly to the Thouless number $g_T$, derived from the microscopic approach, and which is known to be a proper indicator of the localization transition based on the scaling analysis~\cite{Abrahams1979}: $g_T=\langle\gamma_n^{-1}\rangle^{-1}/\langle\delta\omega_n\rangle$ in the situation discussed in the present paper~\cite{Skipetrov2014}, with $\gamma_n$ the inverse lifetimes of the eigenmodes and $\langle\delta\omega_n\rangle$ the average mode spacing. As presented in Figs.~\ref{Fig:condvar}(b) and (c), the conductance $g_C$ collapses from very large values to small ones, in the same region where the Thouless number predicts the transition. A noteworthy difference is that the conductance collapses at the boundary of the transition, but increases again deep in the localized regime, reminiscent of fluctuations that diverge at a phase transition~\cite{Huang1987}. Differently, the Thouless number presents small values all over the localized regime. For this reason, the fluctuations and the conductance may be a more accurate tool to characterize the transition itself, as a scaling analysis requires tuning the system size, requiring a much more extensive study, be it experimental or theoretical.

 Interestingly, the localization area corresponds to the Ioffe-Regel criterion $k\overline{l}<\alpha$ where the scattering mean free path $\overline{l}$ becomes comparable to the wavelength~\cite{Skipetrov2018}. Indeed, accounting for the Lorentz-Lorenz shift into the evaluation of $\overline{l}$ leads to the following critical detuning to meet the criterion~\cite{Kaiser2000,Kaiser2009}:
\begin{equation}
\delta_c=\frac{\rho\lambda^3}{8\pi^2}\pm \frac{1}{2}\sqrt{3\alpha\frac{\rho\lambda^3}{4\pi^2}-1}.\label{eq:IR}
\end{equation}
We find that the threshold $k\overline{l}=0.5$ provides a good approximation of the critical region, confirming that the Ioffe-Regel provides a qualitative criterion on the localization transition in our system.
Monitoring the variance close to the transition (here realized at fixed detuning), we observe a scaling $\sigma_I^2\sim 1/g_C^2$ at low conductance, in agreement with theoretical predictions~\cite{Akkermans2007}. This demonstrates clearly that the fluctuations in the scattered intensity is a suitable observable to monitor the localization transition.

While the access to the full light statistics may be challenging for  experiments, the deviation from Rayleigh law is already captured by the intensity variance $\sigma_I^2$~\cite{Eloy2018}.  Indeed, as can be observed in Fig.\ref{Fig:condvar}(b), the variance increases well above unity at the localization transition. For example, the distribution function presented in Fig.\ref{Fig:Histograms}(b) corresponds to $\sigma_I^2\approx6.8$. We note that capturing the intensity statistics using the dynamical speckle fluctuations, as discussed in Fig.\ref{Fig:setup}, requires monitoring the system over many coherence times. While in the diffuse limit, this time scale depends on the number of scattering events~\cite{Eloy2018}, a dedicated study will be needed to understand how it is altered by cooperative effects, in order to distinguish clearly subradiance from Anderson localisation.

These striking similarities between the results of the scaling analysis and the present ones suggests that the intensity variance is an observable suitable to observe the localization transition, and that the changes in the conductance $g_C$ are indeed associated to that transition. Furthermore, to circumvent the possible role of finite-size effects, we have checked that our results hold both at fixed system size $kL$ varying $N$, and at fixed atom number $N$ varying $kL$. However, the size of the incident beam is critical, as its waist must be significantly smaller than the cloud. Indeed, we did not observe an increase of the fluctuations for an incident plane-wave or a Gaussian beam with a large waist: this observation is consistent with the fact that the radiation of a cloud illuminated by a larger laser beam will present a strong component of single scattering~\cite{Weiss2018}, which does obey Rayleigh statistics. Finally, we have checked that the variance is a self-averaging quantity, i.e., it can be computed either using a fixed azimuth angle and different configurations, or studying the fluctuations over the azimuth angles with a single configuration, leading to the same conclusions.

\begin{figure}
    \centering
    \includegraphics[width=1\linewidth]{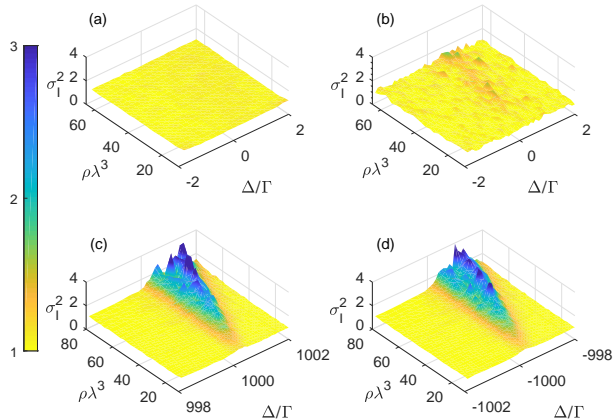} 
    \caption{\label{fig:vect}Intensity variance for the full vectorial model (a) in absence of magnetic field, and (b--d) with a strong magnetic field $\mathbf{B}=B\hat{z}$ for, resp., the $m=0,\ +1$ and $-1$ sublevels. Intensity radiated by an homogeneous cylinder of length $L=21.5/k$, with diameter $L$ as well, which shares its main axis with the laser. (a, c--d) Laser propagating along the $z$--axis with circular polarization $\sigma^\pm$ in the $(x,y)$--plane, with an observation angle $\theta=5\pi/12$, fluctuations computed using $400$ realizations and $51$ values of the azimuth angle. (b) Laser propagating along the $y$--axis, with a linear polarization $\mathbf{E}=E\hat{z}$ parallel to the magnetic field, an observation angle in the $(yz)$--plane at $(\theta,\varphi)=(\pi/12,\pi/2)$, fluctuations  computed using $400$ realizations.}
\end{figure}

Our results hold so far in the scalar light approximation, when polarization effects are neglected. While polarization and near-field terms preclude localization~\cite{Skipetrov2014}, applying a strong magnetic field $B$ restores it~\cite{Skipetrov2015,Skipetrov2018b}, as the atomic system is essentially split into three decoupled scattering subsystems. The latter are associated to different transitions, and get split in energy by an energy $\Delta_B=g_e\mu_B B/\hbar$, with $g_e$ the Land\'e factor of the excited state and $\mu_B$ the Bohr magneton. Turning our analysis to a full vectorial model~\cite{Lehmberg1970}, the study of the light statistics shows that the intensity variance captures the same phenomenology as the eigenvalue analysis~\cite{Skipetrov2018b}: it does not present significant deviations from unity without magnetic field (see Fig.\ref{fig:vect}(a)), yet in presence of a strong magnetic field ($\Delta_B=10^3\Gamma$) the variance increases substantially when addressing the $m=\pm1$ excited atomic states (see Fig.\ref{fig:vect}(c--d)), although the $m=0$ state does not present any signature of the transition (see Fig.\ref{fig:vect}(b)). The only difference that appears between the case of scalar light and that of the $m=\pm1$ transitions in presence of magnetic field is that in the former case, the variance increases only at the transition, and is close to unity in the localized regime far from the transition (see Fig.\ref{Fig:condvar}(a)), whereas it appears to be above unity in all the localized regime in the latter case. Nevertheless, the overall excellent agreement of the deviation of the variance from unity as compared to the eigenvalue analysis, both in the regime considered and in the set of parameters at which it occurs, suggests that light statistics is suitable to detect the localization transition.

In conclusion, we have filled the gap between the macroscopic approach to localization, widely used in experimental works, and the microscopic predictions (eigenvalues analysis), thus demonstrating the equivalence of the two approaches in the context of light scattering by point-like particles. We have shown that the statistics of scattered light is a suitable observable to probe the Anderson localization phase transition for light in 3D, differently from the average intensity. Experimentally, this can be achieved either through configuration averages using atoms that are motionless over the timescale of the measurement, or by studying the time fluctuations of the intensity as the atomic motion makes the system explore various speckle configurations. Interestingly, this approach does not require any finite size scaling analysis. 
We note that the critical density of $20/\lambda^3$ corresponds, for the case of Ytterbium on the $^1S_0\rightarrow ^3P_1$ transition at $\lambda=556nm$, to an atomic density of $\sim 10^{14}atoms/cc$, which can be realized in experiments~\cite{BoschAguilera2018}. 

Part of this work was performed in the framework of the European Training network ColOpt, which is funded by the European Union Horizon 2020 program under the Marie Skodowska-Curie action, grant agreement 72146. R. B. benefited from Grants from São Paulo Research Foundation (FAPESP) (Grants Nos. 2014/01491-0,2015/50422-4 and 2018/01447-2). R. B. and R. K. received support from project CAPES-COFECUB (Ph879-17/CAPES 88887.130197/2017-01). The Titan X Pascal used for this research was donated by the NVIDIA Corporation.

\bibliography{./../../Biblio/BiblioCollectiveScattering}

\end{document}